\documentclass[a4paper,11pt]{article}
\usepackage[a4paper, total={17cm, 25cm}]{geometry}
\usepackage{graphics}
\usepackage{graphicx}
\usepackage{fancyhdr}
\fancypagestyle{plain}{
\fancyhf{} 
\fancyhead[RH]{\thepage}
\fancyfoot[CF]{28th Texas Symposium on Relativistic Astrophysics \\ Geneva, Switzerland -- December 13-18, 2015 \\ \copyright~Commonwealth of Australia (Geoscience Australia) 2016}

}
\begin{document}

\title{\bf Search for the primordial gravitational waves with Very Long Baseline Interferometry}

\author{
Oleg Titov \\ Geoscience Australia, PO Box 378, Canberra, ACT, Australia
\\
S\'ebastien Lambert \\ SYRTE, Observatoire de Paris, PSL Research University, CNRS, \\ Sorbonne Universit\'es, UPMC Univ. Paris 06, LNE
}

\date{February 15, 2016}

\maketitle

\begin{abstract}
Some models of the expanding Universe predict that the astrometric proper motion of distant radio sources embedded in space-time are non-zero as the radial distance from observer to the source grows. Systematic proper motion effects would produce a predictable quadrupole pattern on the sky that could be detected using Very Long Baseline Interferometry (VLBI) technique. This quadrupole pattern can be interpreted either as an anisotropic Hubble expansion, or as a signature of the primordial gravitational waves in the early Universe. We present our analysis of a large set of geodetic VLBI data spanning 1979--2015 to estimate the dipole and quadrupole harmonics in the expansion of the vector field of the proper motions of quasars in the sky. The analysis is repeated for different redshift zones.
\end{abstract}


\section{Introduction}

Very long baseline interferometry (VLBI) measures the differential arrival times of signals from extragalactic radio sources. It is currently the most powerful technique for measuring absolute positions of thousands of radio sources, the orientation and rotation speed of the Earth and ground-based station coordinates, with an accuracy of about 1~cm (or 0.1~mas). VLBI has been extensively used for astrometry and geodesy for about 30 years and, since 1998, is operated by the International VLBI Service (IVS, Schuh \& Behrend 2012). VLBI allows an astrometric precision of $\sim40$~$\mu$as (Fey et al. 2015).

A dipole systematic caused by the galactocentric acceleration of the Solar System Barycentre was detected for the first time in VLBI data by Titov et al. (2011, 2013). The measured amplitude of the aberration drift is in good agreement with the value predicted by Galactic models (e.g., Reid et al. 2009). The quadrupole component presents some systematics that have to be clarified. A major interest of this component is that it can constrain Hubble Constant anisotropy or the amplitude of 
primordial gravitational waves (Kristian \& Sachs 1966; Gwinn et al 1997).

\section{The Galactic Aberration}

The Galactic aberration, or secular aberration drift, is a small proper motion of a few microarc seconds affecting distant bodies induced by the rotation of the Solar system about the Galactic center, which takes about 250~Myr (Kovalevsky 2003). This systematic effect appears as a dipolar deformation of the proper motion field towards the Galactic center ($\alpha=266^{\circ}$, $\delta=-29^{\circ}$) with a magnitude about 6 $\mu$as/yr, corresponding to a 
Solar system acceleration of $3\times10^{-13}$~km/$s^{2}$ in accordance with the equation as follows
\begin{eqnarray} \label{dip}
\Delta\mu_{\alpha}\cos\delta&=&-d_1\sin\alpha+d_2\cos\alpha, \\
\Delta\mu_{\delta}&=&-d_1\cos\alpha\sin\delta-d_2\sin\alpha\sin\delta+d_3\cos\delta,
\end{eqnarray}
where the $d_i$ are the components of the acceleration vector in units of the proper motion, and which corresponds to the degree 1 spheroidal (or electric) development of (e.g., Mignard \& Morando 1990)
\begin{equation} \label{gen}
\vec\mu=\sum_{l,m}\left(a_{l,m}^E\vec Y_{l,m}^E+a_{l,m}^M\vec Y_{l,m}^M\right),
\end{equation}
where $d_1=a_{1,1}^E$, $d_2=a_{1,-1}^E$, and $d_3=a_{1,0}^E$, and $Y_{l,m}^E$ and $Y_{l,m}^M$ the vector spherical harmonics of electric and magnetic types of degree $l$ and order~$m$.

\section{Rotation, Primordial Gravitational Waves, and Anisotropic Expansion of the Universe}

In addition to the aberration distortion, there may also be a small global rotation which can be described by the toroidal (or magnetic) harmonics of degree~1:
\begin{eqnarray} \label{rot}
\Delta\mu_{\alpha}\cos\delta&=&r_1\cos\alpha\sin\delta+r_2\sin\alpha\sin\delta-r_3\cos\delta, \\
\Delta\mu_{\delta}&=&-r_1\sin\alpha+r_2\cos\alpha,
\end{eqnarray}
where the $r_i$ can be expressed in terms of vector spherical harmonics coefficients as $r_1=a_{1,1}^M$, $r_2=a_{1,-1}^M$, and $r_3=a_{1,0}^M$.

Along with the aberration and rotation, more advanced cosmological effects may be detected using the proper motion of distant quasars. In particular, the anisotropic expansion of the Universe would result in the degree 2 vector spherical harmonics of electric type, and the primordial gravitation waves would be an origin of the degree 2 harmonics of electric and magnetic types. To investigate a possible quadrupolar anisotropy of the velocity field, let us give the development of the degree 2 vector spherical harmonics (i.e., $l=2$ in Eq.~(\ref{gen})):
\begin{eqnarray} \label{quad}
\Delta\mu_{\alpha}\cos\delta&=&-(a_{2,2}^{E,\rm Re}\sin 2\alpha-a_{2,2}^{E,\rm Im}\cos 2\alpha)\cos\delta+(a_{2,1}^{E,\rm Re}\sin\alpha-a_{2,1}^{E,\rm Im}\cos\alpha)\sin\delta \nonumber \\
              & &+(a_{2,2}^{M,\rm Re}\sin 2\alpha-a_{2,2}^{M,\rm Im}\cos 2\alpha)\sin\delta\cos\delta+(a_{2,1}^{M,\rm Re}\sin\alpha-a_{2,1}^{M,\rm Im}\cos\alpha)\cos2\delta \nonumber \\
              & &-a_{2,0}^M\sin\delta\cos\delta, \\
\Delta\mu_{\delta}&=&-(a_{2,2}^{E,\rm Re}\cos 2\alpha+a_{2,2}^{E,\rm Im}\sin 2\alpha)\sin\delta\cos\delta-(a_{2,1}^{E,\rm Re}\cos\alpha+a_{2,1}^{E,\rm Im}\sin\alpha)\cos 2\delta \nonumber \\
         & &+(a_{2,2}^{M,\rm Re}\cos 2\alpha+a_{2,2}^{M,\rm Im}\sin 2\alpha)\cos\delta-(a_{2,1}^{M,\rm Re}\cos\alpha+a_{2,1}^{M,\rm Im}\sin\alpha)\sin\delta \nonumber \\
         & &+a_{2,0}^E\sin\delta\cos\delta.
\end{eqnarray}

\begin{figure}[htbp]
\begin{center}
\includegraphics[width=8cm]{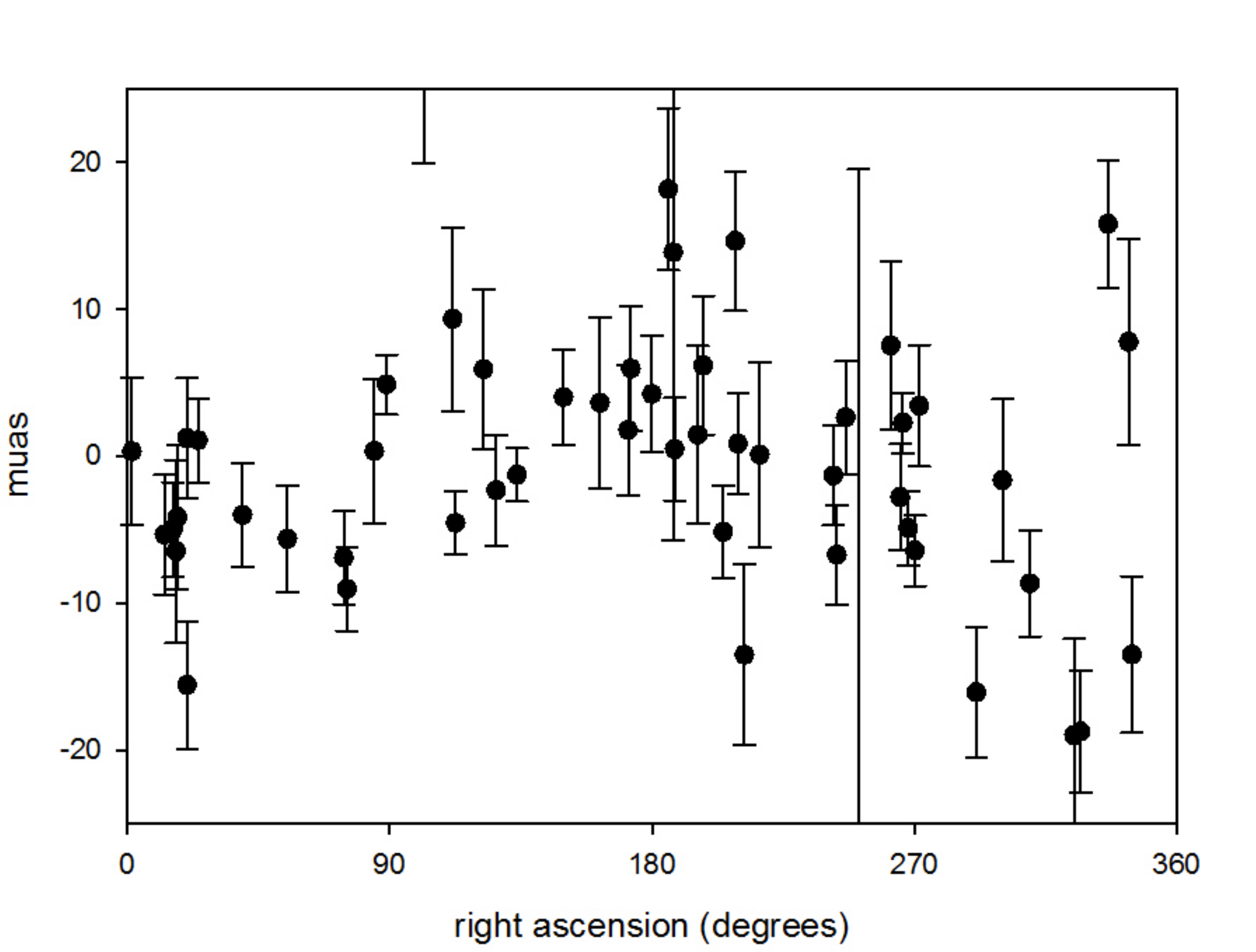}
\includegraphics[width=8.5cm]{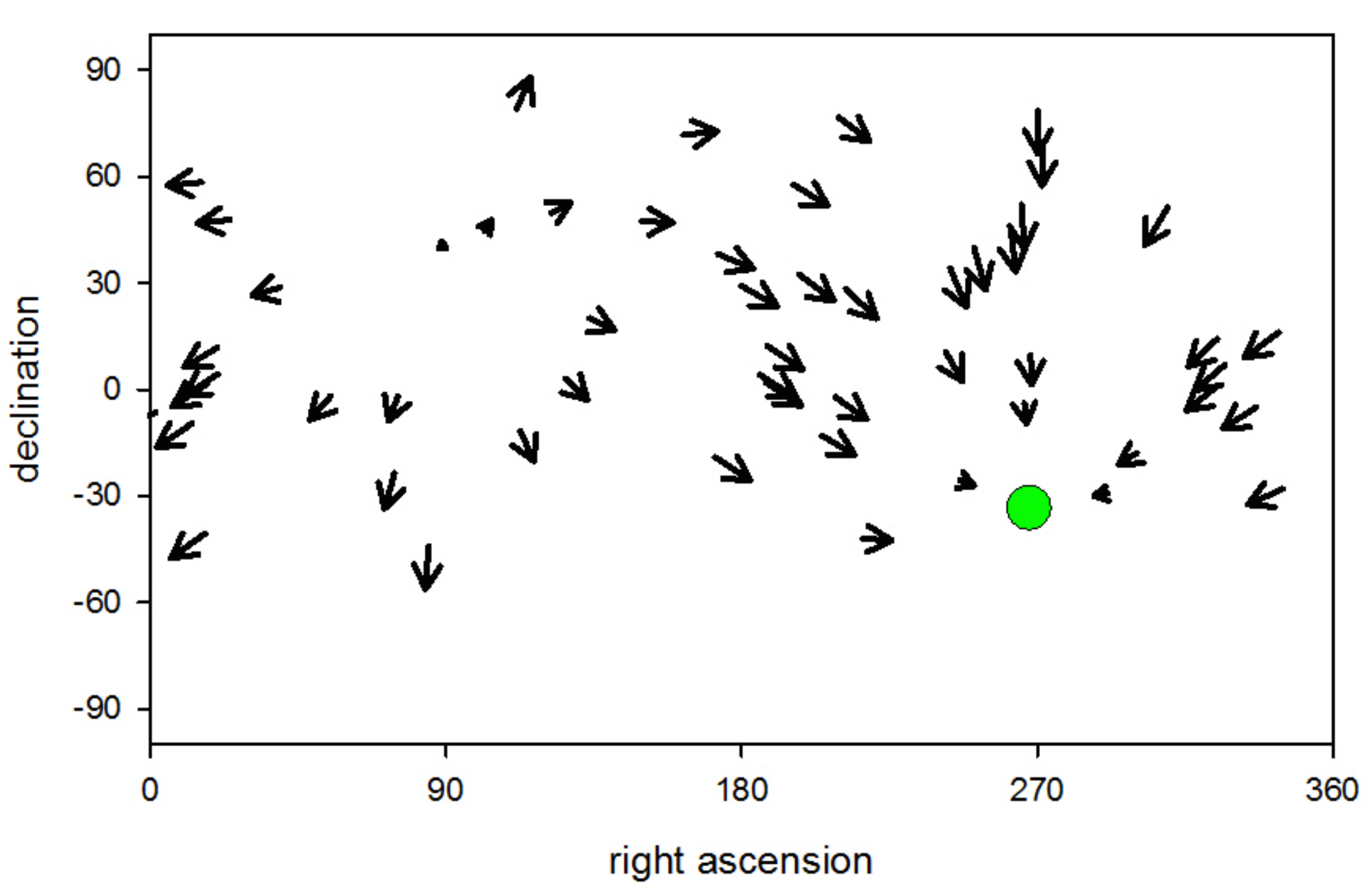}
\end{center}
\caption{The proper motion in right ascension (left) and dipole systematics (right) for the radio sources observed in more than 1000 sessions. The green dot indicates the position of the Galactic center.}
\label{figfit1}
\end{figure}

\begin{table}[htbp]
\caption[]{Parameters estimated from different subsets of radio sources.}
\label{tabfit}
\begin{center}
\begin{tabular}{lccc}
\hline
\hline
\noalign{\smallskip}
 & Dipole only & Dipole + rotation & 16 parameters \\
\noalign{\smallskip}
\hline
\noalign{\smallskip}
\multicolumn{4}{c}{55 radio sources observed in more than 1000 sessions.} \\
\noalign{\smallskip}
\hline
\noalign{\smallskip}
Amplitude ($\mu$as/yr)          & $5.8\pm1.5$ & $5.9\pm1.5$ & $6.3\pm2.1$ \\
$\alpha$ ($^{\circ}$)           & $257\pm19$  & $277\pm20$  & $270\pm21$  \\
$\delta$ ($^{\circ}$)           & $-54\pm14$  & $-49\pm15$  & $-31\pm17$  \\
Rotation ($\mu$as/yr)           &             & $4.4\pm1.6$ & $4.6\pm2.0$ \\
Second harmonics ($\mu$as/yr)   &             &             & $5.1\pm1.8$ \\
\noalign{\smallskip}
\hline
\noalign{\smallskip}
\multicolumn{4}{c}{All 617 radio sources.} \\
\noalign{\smallskip}
\hline
\noalign{\smallskip}
Amplitude ($\mu$as/yr)          & $5.9\pm1.0$ & $6.0\pm1.0$ & $6.4\pm2.1$ \\
$\alpha$ ($^{\circ}$)           & $273\pm13$  & $278\pm14$  & $289\pm13$  \\
$\delta$ ($^{\circ}$)           & $-56\pm9$   & $-54\pm9$   & $-41\pm11$  \\
Rotation ($\mu$as/yr)           &             & $2.2\pm0.8$ & $2.9\pm1.0$ \\
Second harmonics ($\mu$as/yr)   &             &             & $4.4\pm1.1$ \\
\noalign{\smallskip}
\hline
\noalign{\smallskip}
\multicolumn{4}{c}{378 `distant' radio sources with $z>0.9$.} \\
\noalign{\smallskip}
\hline
\noalign{\smallskip}
Amplitude ($\mu$as/yr)          & $7.9\pm1.4$ & $8.1\pm1.4$ & $8.3\pm1.7$ \\
$\alpha$ ($^{\circ}$)           & $267\pm15$  & $277\pm14$  & $297\pm19$  \\
$\delta$ ($^{\circ}$)           & $-59\pm9$   & $-54\pm10$   & $-49\pm12$  \\
Rotation ($\mu$as/yr)           &             & $3.1\pm1.5$ & $3.7\pm1.9$ \\
Second harmonics ($\mu$as/yr)   &             &             & $4.8\pm1.6$ \\
\noalign{\smallskip}
\hline
\noalign{\smallskip}
\multicolumn{4}{c}{256 `close' radio sources with $z<0.9$.} \\
\noalign{\smallskip}
\hline
\noalign{\smallskip}
Amplitude ($\mu$as/yr)          & $3.7\pm1.3$ & $3.5\pm1.4$ & $4.1\pm1.5$ \\
$\alpha$ ($^{\circ}$)           & $285\pm23$  & $284\pm31$  & $283\pm22$  \\
$\delta$ ($^{\circ}$)           & $-44\pm20$   & $-48\pm22$   & $-17\pm23$  \\
Rotation ($\mu$as/yr)           &             & $3.1\pm1.1$ & $3.6\pm1.2$ \\
Second harmonics ($\mu$as/yr)   &             &             & $7.8\pm1.5$ \\
\noalign{\smallskip}
\hline
\end{tabular}
\end{center}
\end{table}

\section{Observations and results}

About 10 million VLBI observations since 1979 were analyzed with the geodetic VLBI analysis 
software Calc/Solve to generate astrometric coordinate time series of about 3800 radio sources. 
Amplitudes and direction of dipole, rotation and second order harmonics are displayed in Table 1 (below)
for various subsets of radio sources. The figures show the proper motion in right ascension 
and dipole systematics for the most observed sources, as well as the electric part of the quadrupole 
systematics for closest sources. 

\begin{figure}[htbp]
\begin{center}
\includegraphics[width=10cm]{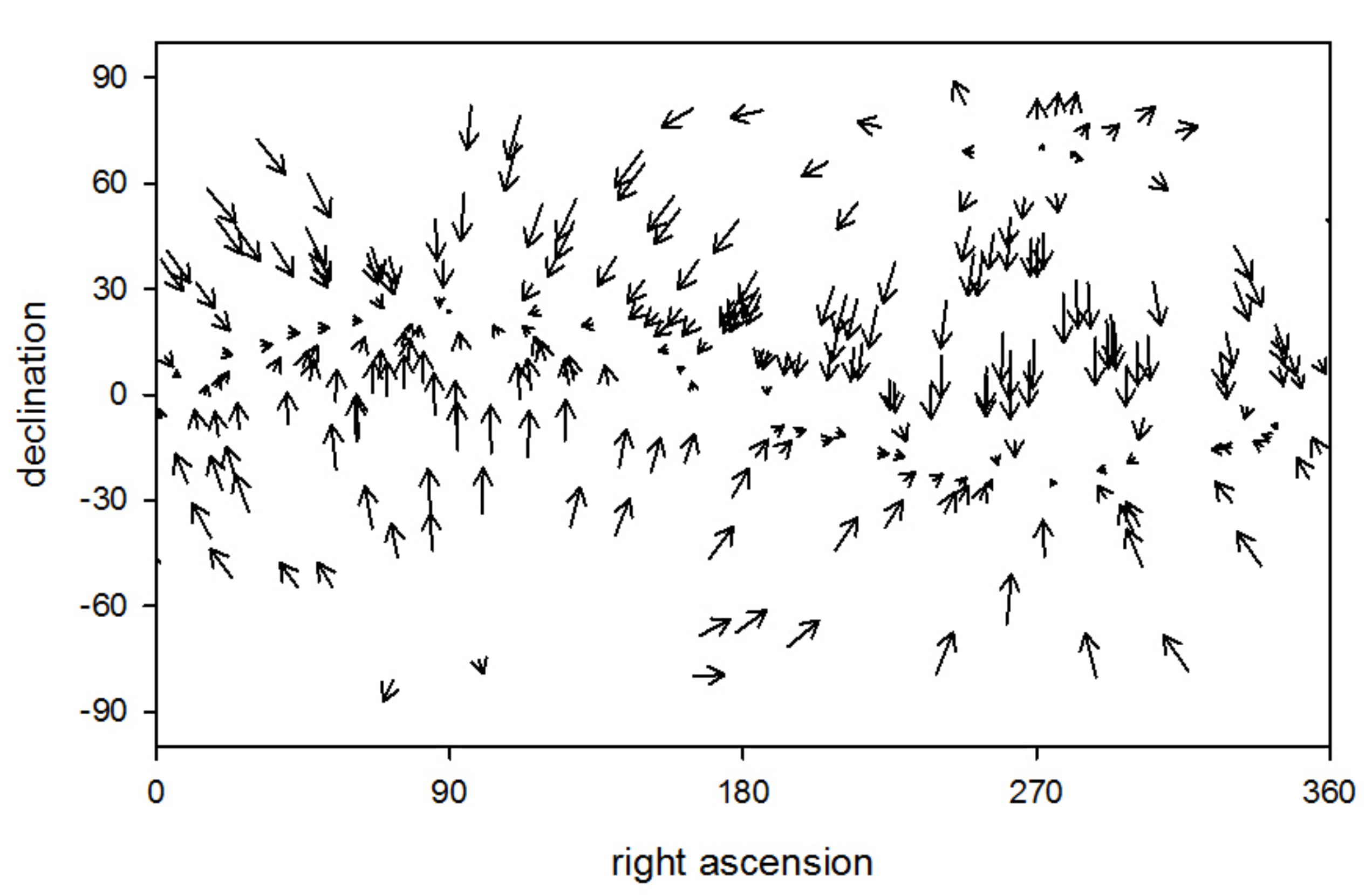}
\end{center}
\caption{The electric part of the quadrupole systematics for closest sources ($z<0.9$).}
\label{figfit2}
\end{figure}

Table 1 displays the magnitude of the dipole, rotation and second order harmonics for different sets
of reference radio sources. As the magnitude estimates of the second order harmonics exceed the 3-$\sigma$ formal 
errors in some cases, they vary from one solution to another. More VLBI data needs to be collected
to obtain more reliable results.

\section{Acknowledgment}

The paper is published with the permission of the CEO, Geoscience Australia.


\end{document}